\documentclass[11pt]{article}
\usepackage{hyperref}

\pdfoutput = 1

\begin{document}

\title{Fluid Mechanics of Everyday Objects}

\author{N. J. Parziale, J. S. Jewell, B. E. Schmidt, J. Rabinovitch, R. Dunne \\
 \\\vspace{6pt} Graduate Aerospace Laboratories \\ 
California Institute of Technology, Pasadena, CA 91125, USA}

\maketitle

\section*{Introduction}
Several videos of common objects are produced to remind us that we are constantly surrounded by a variety of fluid mechanical processes in our everyday lives. The videos are taken using a Phantom v710 camera set at 1000~fps, with an exposure time of 330~ms. A conventional z-type Schlieren setup (as in \cite{ref:Settles2001}) is used with an Osram Diamond Dragon LUW W5AP continuous LED white light source. The movies are played back at 25~fps, which corresponds to a reduction in speed by a factor of 40.

\section*{Part 1: Vortex Ring Production - Water Bottle}
A plain, empty plastic water bottle is seeded with acetone vapor (for density contrast), and is lightly squeezed. The jet exiting from the bottle rolls up and produces an axisymmetric vortex ring, which is quickly followed by a second vortex ring undergoing a leap frog effect. These vortex rings are then followed by a single trailing jet, and start to break down near the end of the clip.

\section*{Part 2: Multiphase Jet - Carbonated Beverage}
A standard 12~oz. can of a generic carbonated beverage is pressurized (through vigorous agitation), and then carefully opened. As the tab initially penetrates the can, it is possible to see the release of an initial pressurized jet, which is then followed by a stronger jet composed of a gas/vapor mixture. Liquid droplets are released from the can and a true multiphase flow develops, with droplets on a variety of different length scales being observed.

\section*{Part 3: Projectile Motion - Cork in Water Bottle}
An empty plastic water bottle is once again seeded with acetone vapor, and a cork is placed in the neck of the bottle. The bottle is quasi-adiabatically compressed with a rubber impactor and projectile motion ensues. 

\section*{Part 4: Flame Propagation - Acetone Vapor Ignition}
A lit match is brought in close proximity to the neck of a plastic water bottle that has been seeded with acetone vapor. At first it is possible to visualize the hot plume from the match itself, and then a spherical flame kernel propagates towards the fuel stream and into the bottle, after which a stream of hot combustion products jets out of the bottle.

\section*{Video}

The video showing these everyday phenomena can be seen at the following URLs:

\begin{itemize}
  \item \href{http://www.its.caltech.edu/~jrabinov/Everyday_Small.mpg}{Link 1: Low Resolution} 
  \item \href{http://www.its.caltech.edu/~jrabinov/Everyday_Large.mp4}{Link 2: High Resolution}
\end{itemize}

High speed Schlieren videos were produced highlighting the fluid mechanics found in everyday objects. This video (entry 102369) was submitted as part of the Gallery of Fluid Motion 2013, which is a showcase of fluid dynamics videos.

\bibliographystyle{plain}

\begin{thebibliography}{99}

\bibitem{ref:Settles2001}
G. S. Settles.
{\it Schlieren and Shadowgraph Techniques}.
Springer Berlin Heidelberg, 2001.

\end{thebibliography}

\end{document}